\documentclass[twocolumn, prd, amssymb, preprintnumbers]{revtex4}
\usepackage{amssymb}
\usepackage{amsmath}
\usepackage[dvips]{graphicx}
\usepackage{epsfig}

\begin{document}  

\preprint{ITFA 2008-38, arXiv:0809.5062v1}
\title{Black Hole Berry Phase}
\author{Jan de Boer, Kyriakos Papadodimas and Erik Verlinde}
\affiliation{ Institute for Theoretical Physics, University of Amsterdam\\
              Valckenierstraat 65, 1018 XE Amsterdam, The Netherlands\\
              J.deBoer, K.Papadodimas, E.P.Verlinde@ uva.nl} 

\bigskip

\date{September 2008}

\begin{abstract}
Supersymmetric black holes are characterized by a large number of
degenerate ground states. We argue that these black holes, like other
quantum mechanical systems with such a degeneracy, are subject to a
phenomenon which is called the geometric or Berry's phase: under
adiabatic variations of the background values of the supergravity
moduli, the quantum microstates of the black hole mix among
themselves. We present a simple example where this mixing is exactly
computable, that of small supersymmetric black holes in 5
dimensions. While in practice it would be extremely difficult to
measure Berry's phase for black holes, it may be interesting to
explore it further from a theoretical point of view, as it provides us
with a way to probe, and to some degree manipulate, the quantum
microstate of the black hole. 

\end{abstract}

\maketitle 

It is believed that ultimately black holes should have a
complete quantum mechanical description. While this goal has not been
achieved for general black holes, string theory has provided us with a
microscopic description of certain kinds of supersymmetric black holes
in terms of the quantum theory living on bound states of D-branes
\cite{Strominger:1996sh}. As evident from their non-vanishing
Bekenstein-Hawking entropy, supersymmetric black holes are
characterized by a large number of degenerate ground states.  The
classical black hole geometry is in a certain sense a
``thermodynamic'' description of this ensemble of microstates.

Seen as a localized compact object, a supersymmetric black hole is
embedded in background flat space which is characterized by a set of
continuous parameters, the moduli of the supergravity theory.  These
will typically enter the quantum D-brane theory describing the black
hole in the form of effective coupling constants. It is then natural
to ask what happens to the microstates of the black hole under
adiabatic changes of the background moduli. The attractor mechanism
\cite{Ferrara:1995ih} guarantees that the entropy of the black hole,
that is the number of ground states of the quantum mechanical system,
does not change under such a variation of the moduli. However this
does not mean that the individual quantum microstates remain invariant
under such a process.

Generally in quantum mechanics this type of questions leads to the
concept of Berry's phase \cite{Berry:1984jv} and its
generalizations. Consider a quantum mechanical system whose
Hamiltonian depends on a set of continuous parameters $\lambda$ taking
values in a parameter space ${\cal M}$. First we assume that the
system has a unique ground state for each value of $\lambda$, which
satisfies
\begin{equation}
  H(\lambda) \psi_0(\lambda) = E_0(\lambda) \psi_0 (\lambda).
\end{equation}
If we adiabatically change the parameters from $\lambda_1$ to
$\lambda_2$ along a path $C$ then the ground state $\psi_0(\lambda_1)$
is transformed into $\psi_0(\lambda_2)$, up to an overall phase, which
turns out to be dependent on the path $C$. This phase is called Berry's
phase \cite{Berry:1984jv} and is is described by a $U(1)$ gauge field
living on the parameter space ${\cal M}$.

More generally, if the system has a set of $n$ degenerate ground
states for all values of $\lambda$ then Berry's phase is promoted to
its nonabelian version studied by Wilczek and Zee
\cite{Wilczek:1984dh}. It is described by a nonabelian $U(n)$ gauge
field on ${\cal M}$. This gauge field can be defined in the following
way. At every point on ${\cal M}$ we have the same Hilbert space
${\mathbb H}$ of the quantum mechanical system, equipped with its
inner product. This defines the flat bundle ${\cal M}\times {\mathbb
  H}$, at least over an open, simply connected domain on ${\cal M}$ which 
does not
include any singular points or points where we have energy level crossings. 
 For every $\lambda$ 
the space of degenerate ground states
${\mathbb H}_0(\lambda)$ is a subspace of ${\mathbb H}$ and forms a
vector bundle over ${\cal M}$, which is a sub-bundle of the flat
bundle ${\cal M}\times {\mathbb H}$. The inner product on this flat
bundle naturally induces a connection $A_i$ on the sub-bundle of
ground states.  For every point $\lambda\in{\cal M}$ we choose an
arbitrary orthonormal basis for the ground states $\psi_a(\lambda),a
 =1,...n$. Then Berry's connection is defined by
 \begin{equation}
   A_{i,ab}= (\psi_a,\partial_i\psi_b)
 \label{berrydef}
 \end{equation}
 where $(\,,\,)$ denotes the Hilbert space inner product. It is easy to 
 see that under a change of basis $\psi_a(\lambda) \rightarrow V_a^b(\lambda)
 \psi_b(\lambda)$ then $A_{i,ab}$ does indeed transform as a connection. We also
 have the curvature of the bundle of ground states
 \begin{equation}
   F_{ij} = \partial_i A_j - \partial_j A_i + i[A_i,A_j].
 \label{curva}
 \end{equation}
 If we start at a point $\lambda\in M$ and move along a closed loop $C$, then
 the ground states will come back to themselves up to a mixing by a
 $U(n)$ holonomy determined by the Wilson line of $A_{i,ab}$ along $C$.

 In the supergravity approximation a supersymmetric black hole is
 characterized by its electromagnetic charges $\Gamma=(q_I,p^I)$ and
 its angular momentum. To fully specify the black hole solution we also
 have to give the background asymptotic values $z^i$ of the scalar
 moduli of supergravity. The ADM mass of the black hole is then fixed
 by the BPS bound in terms of the central charge
 \begin{equation}
   M = |Z(\Gamma,z^i)|
 \end{equation}
 in units where the Planck scale is one. In string theory the charges
 of the black hole correspond to the types of D-branes and the cycles they
 wrap on the internal manifold, and they take values in a charge
 lattice $\Lambda$. 

 For each choice of charges $\Gamma\in\Lambda$ we have a (possibly
 strongly coupled) quantum mechanical system living on the bound state
 of the branes, with an associated Hilbert space ${\mathbb
   H}_{\Gamma}$.  The non-vanishing entropy of supersymmetric black
 holes implies that this system has a large number of degenerate and
 supersymmetric ground states states spanning the subspace ${\mathbb
   H}_{\Gamma,0} \subset {\mathbb H}_{\Gamma}$.  The dimensionality of
 this space is related to the entropy of the black hole as
 \begin{equation}
   S(\Gamma) = \log\left(\text{dim} {\mathbb H}_{\Gamma,0}\right).
 \end{equation}
 In the D-brane description of the black hole, the branes are placed in
 a flat supergravity background characterized by the asymptotic values
 of the moduli $z^i$. Then these values will be seen as coupling
 constants in the quantum system of the branes. If we adiabatically
 change the asymptotic values of the moduli $z^i$, then the subspace of
 ground states ${\mathbb H}_{\Gamma,0}$ will vary inside the full
 Hilbert space ${\mathbb H}_{\Gamma}$. As we explained above, such a
 variation will introduce nontrivial Berry's phase for the microstates
 of the black hole.

 Thus we expect to have the following structure: for each black hole
 with charges $\Gamma\in\Lambda$, we have a $U(N)$ vector bundle over
 the supergravity moduli space ${\cal M}$, whose connection $A_i$
 describes how individual microstates of the black hole mix under
 adiabatic variations of the moduli.  These bundles have very large
 rank since the number of states $N= e^{S(\Gamma)}$ grows exponentially
 with $\Gamma$. If we start at a point $p_1\in {\cal M}$ and
 adiabatically change the moduli along a path $C$ to the final point
 $p_2$ then the black hole microstates will be subject to the $U(N)$
 transformation given by
 \begin{equation}
   U(C) = P \exp(i\oint_C A_i dz^i).
 \label{wilson}
 \end{equation}
 If $C$ is a closed loop starting and ending at the same point $p$ then
 $U(C)$ describes the holonomy of the microstates of the black hole
 under such a variation of the moduli.

 For a general supersymmetric black hole it is not clear how to compute
 the connection $A_i$, since we do not have a precise and controllable
 description of the quantum mechanics of the D-branes. The case where
 the microscopic quantum description is most accessible is that of
 supersymmetric black holes in five dimensions. We start with IIB
 string theory compactified on $K3\times S^1$ and call $R$ the radius
 of the $S^1$. We consider a bound state of D1 and D5 branes wrapped on
 the internal manifold. If the radius $R$ is large compared to the
 other scales, then the bound state has a description in terms of a 1+1
 dimensional superconformal field theory with $(4,4)$ supersymmetry and
 central charge $c=6Q_1 Q_5$. Consider states in the Ramond sector of
 the CFT of the form
 \begin{equation}
   L_0 = 0,\qquad \overline{L}_0 = P.
 \label{d1d5p}
 \end{equation}
 These states lead to a D1/D5/P supersymmetric black hole with a horizon
 of finite area \cite{Strominger:1996sh}.

 In the case $P=0$ the situation is simpler, since both sectors of the
 CFT are in their ground state. In this case it can be shown that
 classically the horizon area is zero, so the object is a small D1/D5
 black hole. While the horizon area classically vanishes, the black
 hole still has entropy of order $4\pi\sqrt{Q_1Q_5}$ which presumably
 becomes manifest when higher order corrections to supergravity are
 included \cite{Dabholkar:2004yr},\cite{Dabholkar:2004dq}. This entropy 
is related to the large number
 of Ramond ground states of the CFT, which represent the quantum
 microstates of the black hole.

 In this system it becomes clear how to compute Berry's phase for the
 black hole microstates: we have to compute how the bundle of Ramond
 ground states varies over the moduli space of the CFT. To be more
 precise, the moduli space of the supergravity theory is locally of the
 form
 \begin{equation}
   {\cal M} = {\mathbb R}^+ \times {SO(5,21) \over SO(5) \times SO(21)}
 \label{modbig}
 \end{equation}
 where the first factor corresponds to the size of the $S^1$ while the
 second factor is roughly the moduli space of IIB compactified on $K3$.
 Once we fix the charges $Q_1,Q_5$ the attractor mechanism fixes a
 submanifold
 \begin{equation}
 {\cal M}_{CFT} \simeq {SO(4,21) \over SO(4)\times SO(21)} 
 \label{modsmall}
 \end{equation}
 inside the second factor of \eqref{modbig}. This space has to 
 be identified with the moduli space of the conformal field theory 
 on the D1/D5 bound state. Given a path $C$ in the space ${\cal M}$ we
 can project it to a path $C'$ on ${\cal M}_{CFT}$ using the attractor
 flow. We conjecture that Berry's phase for the black hole microstates
 along $C$ is the same as Berry's phase for the Ramond ground states of
 the CFT along $C'$. The situation becomes more simple if we take a
 decoupling limit and embed the small black hole in asymptotically AdS$_3$
 space, where the moduli space of supergravity exactly coincides with
 that of the CFT and there is no need of projection on the attractor
submanifold.

Berry's phase for the Ramond ground states of the D1/D5 CFT was
computed in \cite{deBoer:2008ss} for somewhat different reasons. The
main tool for the computation is the $tt^*$ equations
\cite{Cecotti:1991me}. These determine the Berry phase of the bundle
of Ramond ground states of a $(2,2)$ superconformal field theory in
terms of the chiral ring coefficients $C_{ij}^k$. 
More precisely the curvature tensor \eqref{curva} is equal to
\begin{equation}\begin{split}
& F_{ij} =0 \\
& F_{\overline{i}\overline{j}} = 0\\ &
    F_{i\overline{j}} =-
    [C_i,\overline{C_j}]+
    g_{i\overline{j}} \left( 1 - {3\over c}(q+ \overline{q})\right)
\end{split}\label{finaltt}\end{equation}
where $g_{i\overline{j}}$ is the inner product between Ramond ground states,
and $q,\overline{q}$ are their left and right moving R-charges. 
Additionally in $(4,4)$ superconformal theories one can show
\cite{deBoer:2008ss} that the chiral ring coefficients are covariantly
constant over the moduli space
\begin{equation}
  \nabla C_{ij}^k= 0.
\label{ringc}
\end{equation}
Acting with $\nabla$ on \eqref{finaltt} and using \eqref{ringc} we
conclude that Berry's curvature for Ramond ground states in $(4,4)$
theories is covariantly constant 
\begin{equation}
  \nabla F =0.
\end{equation}
Bundles of covariantly constant curvature over symmetric spaces such
as \eqref{modsmall} are called homogeneous bundles and their geometry
is completely fixed by the geometry of the base space ${\cal M}_{CFT}$
in terms of some basic group theoretic data. The black hole states fall
into representations of the $SO(21)$ group. The curvature operator 
characterizing Berry's phase for states falling into such a representation
${\cal R}$ can be written as
\begin{equation}
F = -  \delta_{ab} \Sigma^{\cal R}_{IJ}
\end{equation}
where the indices $a,b$ and $I,J$ refer to the $SO(4)$ and $SO(21)$ factors of the
tangent bundle of \eqref{modsmall} respectively and
$\Sigma^{\cal R}$ is the matrix of the ${\cal R}$ representation of $SO(21)$.
Notice that there is no curvature in the $SO(4)$ factor.
More details about these bundles can be found in \cite{deBoer:2008ss}. 
The main physical conclusion is that Berry's phase for the microstates 
of the D1/D5 black hole is exactly computable.

We would like to emphasize that we expect the same phenomenon for most
other cases of supersymmetric black holes, such as those in four dimensional
${\cal N}=2$ supergravity, even though the computation
of the phase may be very difficult with current technology.

We should also stress that the holonomy that we discuss in this paper
is a local holonomy, appearing along contractible loops on the moduli
space. This is different from global monodromies of black holes and
dyons around singularities. In the case of local holonomy the
spacetime charges of the black hole do not change, only microstates
with the same charges mix, while typically the charges shift under
global monodromies \cite{Seiberg:1994rs}. In our discussion we have
assumed that no singularity or wall of marginal stability is
encountered anywhere along the path $C$ on the moduli space. However
it might be interesting to explore the relation between our local holonomy
and the black hole decay on the walls of marginal stability. 

The fact that there is a nontrivial Berry's phase for the ground states
of the black hole means that in principle we can change and tune the
microstate, to a certain degree, by varying the moduli. Also, we can
set up interference experiments which are sensitive to the internal
microstate of the black hole. These are interesting observations from
a theoretical point of view, but of course given the enormous
degeneracy of the microstates of black holes, such a tuning would be
very difficult in practice. Generically for a large black hole we
expect that the phase will change very rapidly with the moduli. In the
case where all microstates with the same charges do mix then we would
expect that the amount of tuning of the moduli necessary to observe the 
phase would be schematically of the order
\begin{equation}
  {\Delta z  \over z } \sim e^{-S}
\end{equation}
where $S$ is the entropy. So the observation of the phase is exponentially
difficult for large $S$. 
In this sense the Berry phase can be viewed as a subtle example of
quantum black hole hair that is indeed almost impossible to measure 
macroscopically so it does not contradict the no hair theorems. See also 
\cite{Balasubramanian:2006iw} for related discussions.

These limitations are of less importance for
small black holes, where Berry's phase would be easier to observe. 
Notice moreover that in the D1/D5 case the
assumption of generic mixing between microstates is not true, but
instead the holonomy for the microstates is ``block-diagonal''
allowing mixing only between certain subsets of microstates
\cite{deBoer:2008ss}.

Putting aside  practical difficulties, we briefly discuss how at
least in principle we can observe and manipulate the Berry's phase for
supersymmetric black holes.  The most straightforward way to change
the microstate of the black hole is to actually adiabatically change
the values of the moduli of supergravity along a path $C$ on ${\cal
  M}$. This is not possible to do dynamically everywhere in space,
given the infinite volume. However it is sufficient to construct a
very big bubble surrounding the black hole in which the values of the
moduli have the desired values, which may then be changed
adiabatically along a path $C\in {\cal M}$. In principle such a
process is allowed and would induce the change of the black hole
microstate determined by Berry's phase \eqref{wilson} for the path
$C$.

Another way to achieve this is by considering a configuration of
several large D1/D5/P black holes with different charges widely
separated in space.  While this system is not supersymmetric (or
static), if these black holes are placed far enough from one-another
then the system will remain approximately static for a very long time.
The values of the scalar moduli near each of these black holes will
flow to the corresponding attractor values determined by the charges
of the black hole. This multi-black hole configuration creates a
slowly varying nontrivial spatial profile $z^i(x)$ for the
supergravity moduli. Then we bring the probe black hole whose
microstate $|\psi\rangle $ we want to tune and move it around slowly
in this complicated background along the trajectory $x^\mu(\tau)$. The
values of the background moduli that the probe feels will depend on
how close it is from the various other big black holes and hence on
its trajectory. In this way we can effectively induce a rotation of
the internal phase of the probe black hole by
\begin{equation}\begin{split}
    & \qquad \qquad |\psi\rangle \rightarrow U |\psi\rangle, \cr & U =P
    \exp\left(i \int_{\tau_1}^{\tau_2} A_i \partial_\mu z^i
    \dot{x}^\mu d\tau\right).
\end{split}\end{equation}
Similarly we can observe this internal phase by setting up
interference or Aharonov-Bohm type experiments sensitive to the
internal microstate of the black hole. See also \cite{Bowick:1988xh},
\cite{Krauss:1988zc}, \cite{Coleman:1991ku} for related discussions in
the context of quantum black hole hair. Of course for large black holes 
this would be very difficult to measure in practice but possible in 
principle. 

It would be interesting to extend the computation of
\cite{deBoer:2008ss} to D1/D5/P states which correspond to macroscopic
black holes in 5 dimensions and are also related to dyonic black holes in
4d ${\cal N}=4$ supergravity. For this one has to compute the holonomy
over the moduli space ${\cal M}_{CFT}$ for states of the form
\eqref{d1d5p} with $P\neq 0$, so only the left moving sector is in its ground state,
while the right moving can be excited. These states are counted by the
elliptic genus of the symmetric product $K3^N/S_N$ and their
degeneracies are encoded in the Siegel modular form $\Phi_{10}$. It
would be interesting to write down the general equations which
determine Berry's curvature for the states contributing to the
elliptic genus generalizing the $tt^*$ equations \eqref{finaltt},
since these would correspond to microstates of a black hole with a
macroscopic horizon. These states are related by spectral flow to
operators in the NS sector of the form (chiral primary , anything). In
principle using the formalism of \cite{Ranganathan:1993vj} the
curvature for these operators can be computed, but it would be nice to
see whether the BPS condition of the left moving sector simplifies the
computation in any way.

It would also be interesting to study the case of 4 dimensional black
holes in ${\cal N}=2$ supergravity. In that case the attractor
mechanism fixes the vector multiplets. Thus we would expect that the
nontrivial part of the Berry's phase will come from motion on the
hypermultiplet moduli space. In principle, if we could write down the
dual superconformal quantum mechanics we would be able to see how the
hypermultiplets enter the theory as effective coupling constants and
compute Berry's phase. Hopefully it might be possible to express this
curvature in terms of BPS quantities without having to develop a full
understanding of the black hole quantum mechanics. It is also
conceivable that such a computation can be done using properties of
the MSW (0,4) CFT.

Finally let us note that the Ramond microstates of the D1/D5 system
are related to the supergravity fuzzball solutions found by Lunin and
Mathur \cite{Lunin:2001jy} and also \cite{Kanitscheider:2007wq}. It
would be nice to see if the holonomy that we computed can also be
reconstructed from the supergravity point of view by studying the
variation of the solutions under adiabatic changes of the moduli. This
set of solutions constitutes the (supersymmetric) phase space of a
Hamiltonian system defined by classical supergravity. This system
depends on continuous parameters, the asymptotic values of the
moduli. The analogue of Berry's phase for classical Hamiltonian
systems is given by the so called Hannay angles \cite{hannay} and
their generalizations. It might be interesting to perform this
classical computation and compare to the CFT results.  For such a
comparison it is important to take into account that the classical
solutions correspond to coherent superpositions of Ramond ground
states \cite{Alday:2006nd}. We leave this analysis for future work.

Other systems in string theory where Berry's phase appears have
been studied recently \cite{Pedder:2007wp},\cite{Pedder:2008je},
\cite{Sonner:2008be}.
 
\centerline{\bf Acknowledgments}

We would like to thank F. Denef, N. Iizuka, M. Shigemori, S. El-Showk,
K. Skenderis, M. Taylor, D. Tong for valuable discussions and
J. Manschot for collaboration on related work \cite{deBoer:2008ss}.
K.P. would like to thank the ``Monsoon Workshop on string theory'' at
TIFR, Mumbai for warm hospitality during the completion of this
work. This work was partly supported by the Foundation of Fundamental
Research on Matter (FOM).

\end{document}